\documentclass[10pt]{article}

\usepackage{cite}
\usepackage[labelfont=bf,labelsep=period,justification=raggedright]{caption}

\makeatletter
\renewcommand{\@biblabel}[1]{\quad#1.}
\makeatother

\date{}

\begin{document}

\begin{flushleft} {\Large \textbf{Predictability of evolutionary
      trajectories in fitness landscapes} }

Alexander E.~Lobkovsky$^{1}$, 
Yuri I.~Wolf$^{1}$,
Eugene V.~Koonin$^{1,\ast}$
\\

\bf{1} National Center for Biotechnology Information, National Library
of Medicine, National Institutes of Health, Bethesda, MD 20894
\\

$\ast$ E-mail: \texttt{koonin@ncbi.nlm.nih.gov}
\end{flushleft}

\section*{Abstract}
Experimental studies on enzyme evolution show that only a small
fraction of all possible mutation trajectories are accessible to
evolution.  However, these experiments deal with individual enzymes
and explore a tiny part of the fitness landscape.  We report an
exhaustive analysis of fitness landscapes constructed with an
off-lattice model of protein folding where fitness is equated with
robustness to misfolding.  This model mimics the essential features of
the interactions between amino acids, is consistent with the key
paradigms of protein folding and reproduces the universal distribution
of evolutionary rates among orthologous proteins.  We introduce mean
path divergence as a quantitative measure of the degree to which the
starting and ending points determine the path of evolution in fitness
landscapes.  Global measures of landscape roughness are good
predictors of path divergence in all studied landscapes: the mean path
divergence is greater in smooth landscapes than in rough ones.  The
model-derived and experimental landscapes are significantly smoother
than random landscapes and resemble additive landscapes perturbed with
moderate amounts of noise; thus, these landscapes are substantially
robust to mutation.  The model landscapes show a deficit of suboptimal
peaks even compared with noisy additive landscapes with similar
overall roughness.  We suggest that smoothness and the substantial
deficit of peaks in the fitness landscapes of protein evolution are
fundamental consequences of the physics of protein folding.

\section*{Author Summary}

Is evolution deterministic, hence predictable, or stochastic, that is
unpredictable?  What would happen if one could ``replay the tape of
evolution:'' will the outcomes of evolution be completely different or
evolution is so constrained that history will be repeated?  Arguably,
these questions are among the most intriguing and most difficult in
evolutionary biology.  In other words, the predictability of evolution
depends on the fraction of the trajectories on fitness landscapes that
are accessible for evolutionary exploration.  Because direct
experimental investigation of fitness landscapes is technically
challenging, the available studies only explore a minuscule portion of
the landscape for individual enzymes.  We therefore sought to
investigate the topography of fitness landscapes within the framework
of a previously developed model of protein folding and evolution where
fitness is equated with robustness to misfolding.  We show that
model-derived and experimental landscapes are significantly smoother
than random landscapes and resemble moderately perturbed additive
landscapes; thus, these landscapes are substantially robust to
mutation.  The model landscapes show a deficit of suboptimal peaks
even compared with noisy additive landscapes with similar overall
roughness.  Thus, the smoothness and substantial deficit of peaks in
fitness landscapes of protein evolution could be fundamental
consequences of the physics of protein folding.

\section*{Introduction}
One of the most intriguing questions in evolutionary biology is: to
what extent evolution is deterministic and to what extent it is
stochastic and hence unpredictable?  In other words, what happens if
``the tape of evolution is replayed:'' are we going to see completely
different outcomes or the constraints are so strong that history will
be repeated
\cite{gould97:_full_house,conway09:_predict_evol,morris10:_evol_predict,koonin10:_constr_plasticity}?
If evolution is envisaged as movement of a population across a fitness
landscape, the question can be reworded more specifically: among the
numerous trajectories connecting any two points on the landscape, what
fraction is accessible to evolution?  Until recently, these remained
purely theoretical questions as experimental study of fitness
landscapes in the actual sequence space was impractical, due both to
the technical difficulty of producing and assaying numerous expressed
sequence variants and to the more fundamental problem of defining an
adequate quantitative measure of fitness.  However, recent
experimental studies of fitness landscapes could potentially shed
light on the problem of evolutionary path predictability.

The most thoroughly characterized feature of empirical fitness
landscapes is the structure near a peak.  In experiments that examine
the peak structure, a high fitness sequence is typically subjected to
either random mutations or an exhaustive set of mutations at a small
number of important sites.  The resulting library of mutants is then
assayed to measure a proxy of fitness
\cite{elena97:_Ecoli_225,parera09:_epist_hiv_proteas,domingo-calap09:_random_mutat_bactphage,miller06:_IMDH,lunzer05:_IMDH_NAD}.
Significant sign epistasis (a situation in which the fitness effect of
a particular mutation can be either positive or negative depending on
the genetic context) near the peak has been typically observed whereas
local deviations from the additive model have been found to be
uncorrelated with the genetic context and derived from a nearly normal
distribution
\cite{beerenwinkel07:_epist_geometric,martin07:_epistas_model,lunzer10:_pervas_crypt_epist,weinreich05:_sign_epistas}.
Because these studies characterize only a small region of the
landscape, they cannot be used to address the question of path
predictability.

Another broad class of experiments probes the evolutionary
trajectories from low to high fitness.  Usually, in such experiments, a
random peptide is subjected to repeated rounds of random mutagenesis
and purifying selection
\cite{miller06:_IMDH,voigt00:_directed_evolution,bershtein06:_random_mut_fit_betalactam,tracewell09:_direc_enzym_evolut,romero09:_fit_land_direc_evolut}.
During this process fitness grows with each generation and eventually
stagnates at a suboptimal plateau.  The characteristics of the fitness
growth as well as the dependence of the plateau height on the library
size can be used to classify landscapes
\cite{kryazhimskiy09:_dynam_adapt_correl_lands}.  A quantitative
comparison to the $NK$ model of random epistatic landscapes ($N$ is
the number of sites in an evolving sequence and $K$ is the number of
sites that affect the fitness contribution of a particular site
through epistatic interactions) can even yield quantitative estimates
of $N$ and $K$
\cite{kauffman89:_nk_model_rugged,hayashi06:_infectivity_rugged_fitnes}.
The directed evolution studies explore the evolutionarily accessible
portion of the landscape and could in principle be used to shed light
on the question of path predictability.  However, the inaccessible
regions of the landscape remain unexplored and the volume of data at
this point is insufficient to obtain quantitative conclusions
regarding path predictability.

A different type of landscapes has been explored in various microarray
experiments where protein-DNA(RNA) binding affinity serves as the
proxy for fitness
\cite{carlson10:_specif_lands_of_dna_bindin,knight09:_array_based_evolut_of_dna}.
These experiments produce vast, densely sampled landscapes.  A
comparison with a sophisticated Landscape State Machine model of a
correlated fitness landscapes yields estimates of the model parameters
\cite{rowe10:_LSM_fitting,rowe10:_compl_dna_prot_aff_land}.  The DNA
binding landscapes, in principle, contain the information required for
the analysis of path statistics, and could be a valuable resource
for advancing the understanding of evolutionary path predictability.

Empirical studies that exhaustively sample a region of the fitness
landscape allow one to actually assess the accessibility of the entire
set of theoretically possible evolutionary trajectories in a
particular (small) area of the fitness landscape.  For example, all
mutational paths between two states of an enzyme, e.g., the transition
from an antibiotic-sensitive to an antibiotic resistant form of
$\beta$-lactamase
\cite{weinreich06:_5point_beta_lactam,lozovsky09:_malaria_pyrimeth,novais10:_evolut_traj_beta_lactam}
or the transition between different specificities of sesquiterpene
synthase \cite{omaille08:_sesquiterpene} have been explored.  The
results of these experiments, which out of necessity explore only
short mutational paths of $<10$ amino acid replacements, suggest that
there is a substantial deterministic component to protein evolution:
only a small fraction of the possible paths are accessible for
evolution
\cite{weinreich06:_5point_beta_lactam,poelwijk07:_empir_fitnes_lands,kogenaru09:_evolt_path_lands_recon,dawid10:_mult_peaks_recip_sign}.


Theoretical analyses of available empirical fitness data reveal a
tight link between genetic and molecular interactions which are
responsible for the landscape ruggedness and ubiquitous sign epistasis
\cite{weinreich05:_sign_epistas,poelwijk11:_epist_peaks}.  The
emerging quantitative analysis of fitness landscapes can shed light on
some of the most fundamental aspects of evolution but the
interpretation of the currently available experimental results
requires utmost caution as only a minuscule part of the sequence space
can be explored, and that only for a few more or less arbitrarily
selected experimental systems.

Here we focus on the question of the predictability of mutational
paths which is intimately tied to the ruggedness/smoothness of the
fitness landscapes.  The study of random landscapes of low
dimensionality revealed an intuitively plausible negative correlation
between the roughness of a landscape and the availability of pathways
of monotonic fitness \cite{carneiro10:_colloq_paper}.  In the same
study, Carneiro and Hartl showed that experimentally characterized
landscapes are significantly smoother than their permuted counterparts
and exhibit greater peak accessibility
\cite{carneiro10:_colloq_paper}.

To gain insights into the structure of the fitness landscapes of
protein evolution and in particular the accessibility of mutational
paths we used a previously developed simple model of protein folding
and evolution \cite{lobkovsky10:_univer_distr}.  The key assumption of
this model, which is based on the concept of misfolding-driven
evolution of proteins
\cite{drummond08:_MIM,drummond09:_evol_err_prot_synth,wolf10:_relat_contr_struc_func},
is that the fitness of model proteins is determined solely by the
number of misfolded copies that are produced before the required
abundance of the correctly folded protein is reached.  We have
previously shown that this model accurately reproduces the shape of
the universal distribution of the evolutionary rates among orthologous
protein-coding genes along with the dependencies of the evolutionary
rate on protein abundance and effective population size
\cite{lobkovsky10:_univer_distr}.  These results appear to suggest
that our folding model (described in detail the Methods section) is
sufficiently rich to reproduce some of the salient aspects of
evolution.  The model is also simple enough to allow exhaustive
exploration of the fitness landscapes, which prompted us to directly
address the problem of evolutionary path predictability.

We build on the efforts of Carneiro and Hartl
\cite{carneiro10:_colloq_paper} who examined the statistics of
evolutionary trajectories.  Although counting monotonic fitness paths
reveals important features of the landscapes, we argue that reliable
retrodiction of the evolutionary past is possible (i.e., evolution is
quasi-deterministic) only when the available monotonic paths are
similar to each other in a quantifiable way.  We therefore propose a
measure of path divergence to quantify the difference between the
available monotonic paths.  Our aims are to investigate the structure
of the fitness landscapes of protein evolution and to elucidate the
connection between the roughness of landscapes and the predictability
of mutational trajectories.  We analyze three classes of fitness
landscapes: landscapes in which fitness is derived from the folding
robustness of model polymers; additive random landscapes perturbed by
noise; and experimental landscapes derived from the combinatorial
mutation analysis of drug resistance and enzymatic activity.  We show
that all three classes of landscapes are markedly smoother than their
randomly permuted counterparts and all exhibit a similar qualitative
connection between roughness and path predictability.  However, at the
same level of path predictability, the folding landscapes have
substantially fewer fitness peaks.  Given that the statistical
properties of the model landscapes can be directly traced to the
constraints imposed by the energetics and kinetics of a folding
heterpolymer, we hypothesize that the relative smoothness and the
suppression of suboptimal peaks in fitness landscapes of protein
evolution are fundamental consequences of protein folding physics.

\section*{Results}
\label{sec:results-discussion}

\subsection*{Quantitative characterization of fitness landscapes}
\label{sec:quant-char-fitn}

Carneiro and Hartl compared small random landscapes to several
empirical fitness landscapes using deviation from additivity as a
measure of roughness \cite{carneiro10:_colloq_paper}.  They found that
empirical landscapes were significantly smoother than their random
counterparts and that the degree of smoothness was correlated with the
number of monotonic paths to the main summit.  Deviation from
additivity of a landscape is computed by fitting an additive model in
which the fitness of each sequence is different from the peak fitness
by the sum of contributions of the substitutions that differentiate it
from the peak sequence.  The negative fitness contributions of the
substitutions to the peak fitness are adjusted to minimize the sum $S$
of squares of the differences between the actual fitnesses in the
landscape and the fitnesses predicted by the additive model.
Deviation from additivity is defined as $\sqrt{S/L}$, where $L$ is the
number of points in the landscape.

Because roughness of a multidimensional landscape with variable degree
connectivity is not an intuitive concept, we introduce three
additional quantitative measures to probe alternative facets of the
concept of roughness.  First, local roughness is the root mean squared
difference between the fitness of a point and its neighbors, averaged
over the entire landscape.  As defined, local roughness conflates the
measures of roughness and ``steepness.''  For example, a globally
smooth landscape, in which fitness depends only on the distance from
the peak, will have a non-zero local roughness.  However, because
there is a large number of directions that change the distance from
the peak by one, the local roughness of a globally smooth landscape
will be vanishingly small.  In addition, our landscapes tend to be
globally flat--so that the average decrease in fitness due to a single
mutation step away from the main peak is much smaller than the local
fitness variability--everywhere except a small region around the main
peak (see Fig.\ \ref{fig:7}).  Therefore, the landscape-average local
roughness in our case is a true measure of the local fitness
variability.

Second, the fraction of peaks is the number of points with no fitter
neighbors divided by the total number of points in the landscape.  A
strictly additive landscape has a single peak
\cite{kogenaru09:_evolt_path_lands_recon} whereas the peak fraction in
landscapes derived from the folding model as well as the corresponding
randomized landscapes depends on the method of landscape construction,
alphabet size and sequence length.

Third, the roughness of a landscape can be assessed by identifying its
tree component.  The tree component is the set of all nodes with no
more than one neighbor of higher fitness.  Thus, the tree component
includes peaks and plateaus.  Monotonic fitness paths along the tree
component form a single or several disjoint tree structures without
loops.  In the limit of high selection pressure, a mutational
trajectory that finds itself on the tree component has a single path
to the nearest peak or plateau, i.e.\ evolution on the tree component
is completely deterministic.  We use the mean distance to the tree
component, i.e.\ the distance to the tree component averaged over the
landscape, as a measure of roughness.  In a fully additive landscape,
only the peak sequence and its immediate neighbors belong to the tree
component and therefore the mean distance to the tree component is a
measure of the diameter of an additive landscape (which, for example,
could be defined as the maximum pairwise distance between points on
the landscape).  Kauffman and Levin have shown that in a large class
of correlated random landscapes, the mean distance to the tree
component grows only logarithmically with the number of points in the
landscape \cite{kauffman89:_nk_model_rugged}.

We utilize two quantitative measures of the predictability of
evolutionary trajectories.  First is fraction of monotonic paths to
the main peak $F_m$ which is computed by counting the number $n_i$ of
simple (without reverse substitutions or multiple substitutions at the
same site) monotonic paths to the main peak from each point $i$ on the
landscape, dividing it by the total number of simple paths $h_i!$
(where $h_i$ is the Hamming distance from point $i$ to the peak), and
averaging over the landscape via
\begin{equation}
  \label{eq:7}
  F_m = \frac{1}{L}\sum_i \frac{n_i}{h_i!},
\end{equation}
where $L$ is the number of points in the landscape and the sum
excludes the main peak.  The monotonic path fraction $F_m$ measures
the scarcity of accessible evolutionary paths when selection is
strong.  When the monotonic path fraction is small, evolution is more
constrained.

Second, the mean path divergence, is a fine-grained measure of
evolutionary (un)predictability.  We first define the divergence
$d(p_1,p_2)$ of a pair of paths $p_1$ and $p_2$, as the average of the
shortest Hamming distances from each point on one path to the other
path.  Suppose that we have a way of generating stochastic
evolutionary paths.  The outcome of a large number of evolutionary
dynamics simulations is a collection of paths with their associated
probabilities of occurrence.  In general, the probability of occurrence
of an evolutionary path is proportional to the product of fixation
probabilities of its constituent mutation steps.  Given a bundle of
paths with the same starting and ending points, we define its mean
path divergence to be
\begin{equation}
  \label{eq:1}
  \bar d = \sum_{p_1 \neq p_2}  d(p_1, p_2) \, O(p_1) \,
  O(p_2),
\end{equation}
where $O(p)$ is the probability of occurrence of path $p$ in the
ensemble.  In other words, if two paths were drawn from the bundle at
random with probabilities proportional to $O(p)$, their expected
divergence would be $\bar d$.  Alternatively, if we were to fix one
path to be the most likely path in the bundle and to select the second
path at random with probability proportional to $O(p)$, the divergence
would be proportional to $\bar d$ as well.

In an additive landscape, the mutational trajectory is maximally
ambiguous.  As every substitution that brings the sequence closer to
the peak increases fitness, substitutions can occur in any order and
all shortest mutational trajectories to the peak--without reverse
substitutions or multiple substitutions at the same site--are
monotonic in fitness.  In the strong selection limit of our model
defined below, all monotonic trajectories have roughly the same
probability of occurrence, so the mutational path cannot be predicted.

The mean path divergence is a better measure of the predictability of
evolutionary trajectories than the number or fraction of accessible
paths.  Even when only a small fraction of paths are monotonic in
fitness, these paths could potentially be quite different, perhaps
randomly scattered over the landscape.  In such a case, prediction of
the evolutionary trajectory would be inaccurate despite the scarcity
of accessible paths which will be reflected in a high value of path
divergence.

Equation (\ref{eq:1}) introduces the mean path divergence of a bundle
of paths with the same starting and ending points.  The landscape-wide
mean path divergence is measured by constructing representative path
bundles with all possible [start, peak] pairs including suboptimal
peaks as trajectory termination points.  Path divergence is averaged
over all bundles with the starting and ending points separated by the
same Hamming distance.  To construct the path bundles, we employed a
low mutation rate model in which the attempted substitutions are
either eliminated or fixed in the population before the next mutation
attempt occurs.

We invoke the misfolding-cost hypothesis to assign a fitness to a
sequence that folds with probability $P$ to a particular structure.
To produce an abundance $A$ of correctly folded copies, an average of
$A(1 - P)/P$ of misfolded copies are produced.  The ``fitness'' of a
sequence should be a monotonically decreasing function of the cost
incurred by the misfolded proteins.  Previously we showed that
qualitative conclusions drawn from the average population dynamics on
the fitness landscape did not depend on the precise functional
relationship between the number of misfolded copies and fitness
\cite{lobkovsky10:_univer_distr}.  We use simply the negative of the
number of misfolded copies and assign a fitness $w = -A/P$, to a
sequence whose probability of folding to the reference structure is
$P$.  Because the exact population dynamics model is not important, we
use diploid population dynamics in the low mutation rate limit.
Therefore, the probability of fixation of a mutant $j$ in the
background of $i$ is given by
\begin{equation}
  \label{eq:2}
  \pi(i\rightarrow j) = \frac{1 - e^{-2(w_j - w_i)}}{1 - e^{-4
      N_e (w_j - w_i)}},
\end{equation}
where $N_e$ is the effective population size
\cite{kimura62:_prob_fix_mut_pop} which in all simulations was fixed
at $N_e = 10,000$.  The required abundance $A$ is a measure of the
strength of selection.  In the limit of large $A$, the probability of
fixation of a beneficial mutation is unity whereas neutral and
deleterious mutations are never fixed.  In this limit, all uphill
steps in the fitness landscape are equally likely and all monotonic
uphill trajectories have equal evolutionary significance.

In the analysis that follows, we study the association between
landscape roughness and path predictability for the folding landscapes
and their randomized (also referred to as permuted or scrambled)
versions.  In the scrambled landscapes, the topology (i.e.\
connectivity) of the landscape is preserved but the fitness values are
randomly shuffled.  We also compare the roughness and path
predictability characteristics of the model and the experimental
landscapes for $\beta$-lactamase
\cite{weinreich06:_5point_beta_lactam} and sesquiterpene synthase
\cite{omaille08:_sesquiterpene} to those for noisy additive landscapes
with a continuously tunable amount of roughness.

\subsection*{Evolutionary path predictability in fitness landscapes}

\subsubsection*{Deviation from additivity, local roughness, peak
  fraction, and monotonic paths}

We first establish that the folding and the experimental landscapes
are significantly different from their randomly permuted counterparts.
The deviation from additivity of the folding landscapes is typically
several standard deviations below the mean of their scrambled
counterparts.  Although the additivity hypothesis accounts for less
than 40\% of the fitness variability (computed by comparing the sum of
the squares of the fitnesses in the landscape to the sum of the
squares of the residuals of the additive fitness model fit) in all but
one of the folding landscapes, the deviation from additivity of the
permuted landscapes is substantially greater (Fig.\ \ref{fig:1}A).
The experimental landscapes follow the same pattern, in agreement with
the earlier findings of Carneiro and Hartl
\cite{carneiro10:_colloq_paper}.  Furthermore, both in the folding and
in the experimental landscapes, the fraction of monotonic paths to the
main peak is several standard deviations greater than in the
respective scrambled landscapes (Fig.\ \ref{fig:1}B).  An even more
striking disparity exists between the fraction of peaks in the folding
landscapes and their permuted versions: the folding landscapes contain
at least an order of magnitude fewer peaks than their scrambled
counterparts; the experimental landscapes resemble the folding
landscapes more closely than their own randomized versions
(Fig.~\ref{fig:1}C).

To further characterize the deviation of the folding and experimental
landscapes from their permuted counterparts, each landscape metric was
measured and the mean and standard deviation were computed among 100
randomly permuted landscapes.  We then compute the Z-score (deviation
from the mean measured in the units of the standard deviation) of the
original non-permuted landscape compared to the ensemble of the
permuted landscapes.  This Z-score shows how much more correlated the
original landscape is, as measured by the chosen characteristic,
compared to its scrambled counterparts (Figure \ref{fig:2}).  Notably,
despite the considerable scatter of the Z-score values for the folding
landscapes, they all showed extremely large difference (mean Z-score
greater than 20 standard deviations) from the scrambled landscapes for
all measures, with the sole exception of the monotonic path fraction
(Figure \ref{fig:2}).  The two experimental landscapes also
significantly differed from the scrambled landscapes albeit less so
than the folding landscapes, again with the exception of the monotonic
path fraction in which case the two classes of landscapes had similar
Z-scores (Figure \ref{fig:2}).

Aside from the significant correlation (Pearson $\rho = -0.68$)
between peak fraction and mean distance to the tree component, there
was little or no correlation between the four measures of landscape
roughness (Fig.\ \ref{fig:3}).  Roughness of landscapes of high and
variable dimensionality is impossible to capture by a single quantity.
Therefore, the different measures seam to reveal distinct aspects of
landscape architecture.  The strong negative correlation between the
peak fraction and mean distance to the tree component is due to the
fact that each peak spawns a distinct subset of the tree component.
The higher the density of peaks on the landscape, the larger fraction
of the landscape that is covered by the tree component.  Therefore the
average distance to the tree component declines with the increasing
density of peaks.

\subsubsection*{Path divergence}
\label{sec:path-divergence}

Starting from a random non-peak sequence in the landscape, we
introduced random mutations and accepted or rejected them according to
equation (\ref{eq:2}) until the trajectory arrived at a fitness peak.
This procedure was repeated a large number of times, and path bundles
were constructed for all pairs of starting and ending sequences.  Then
the mean path divergence was computed for each path bundle using
equation (\ref{eq:1}) and averaged over all bundles for which starting
and ending points were separated by the same Hamming distance.  When
selection is weak, all mutations which do not result in a sequence
with zero folding probability are accepted.  Thus, evolution is a
random walk on the landscape and the statistical properties of
evolutionary trajectories are fully determined by the topology of the
landscape (i.e. the connectivity of each node).  Conversely, in the
strong selection limit, only mutations that increase fitness are
fixed.  The mean path divergence varies smoothly between the two
limits (Fig.\ \ref{fig:4}) and saturates at high selection pressure.
In our analysis, we focus on the strong selection limit plateau.  In
the weak selection limit, the diversity of trajectories stems solely
from the number of neighbors of each point; by contrast, in the strong
selection limit, the statistics of the monotonic trajectories depend
on the roughness of the landscape.  Thus, the weak selection limit
probes only the topology of the landscape whereas the strong selection
limit also exposes its topography which appears to be critical for
assessing predictability of evolution.

\subsubsection*{Predictors and correlates of path divergence and
  monotonic path fraction}
\label{sec:pred-corr-path}

All four measures of landscape roughness can serve as predictors of
path divergence and monotonic path fraction to some degree (Fig.\
\ref{fig:5}), in agreement with the notion that each of these measures
reflects salient properties of fitness landscapes.  The properties of
the folding and empirical landscapes are consistent with those of
additive landscapes that were perturbed by a moderate amount of noise
(see Methods for details).  A striking exception is the dearth of
peaks and monotonic paths in folding landscapes all other
characteristics being similar.  Deviation from additivity and fraction
of peaks are negatively correlated with path divergence.  This
relationship captures the intuitive notion that in rough landscapes
there are fewer accessible evolutionary paths than in smooth
landscapes, and furthermore, in rough landscapes, even those paths
that are accessible show the tendency to aggregate within small areas
on the landscape.  Indeed, in both the folding model-derived
landscapes and the experimental landscapes, the mean path divergence
for all Hamming distances between the starting and ending points was
dramatically greater than in scrambled landscapes (Fig.\ \ref{fig:6}).
Interpreting these findings in terms closer to biology, the fitness
landscapes derived from the model as well as experimental landscapes
show greater robustness to mutations than random landscapes: a random
mutation in a model-derived or experimental fitness landscape is more
likely than expected for random landscapes to have no deleterious
effect, leading to another monotonic path to the peak.  Consequently,
evolution on the model-derived and experimental landscapes is less
predictable (deterministic) than it would be on uncorrelated random
landscapes.

In contrast to deviation from additivity, the mean distance to the
tree component is positively correlated with path divergence.  When
the tree component comprises a large fraction of the landscape, the
mean distance to the nearest tree branch is small.  Consequently, the
path divergence is reduced as the paths that reach the tree component
do not deviate from each other from that point onward.  By the same
token, when the tree component is large, there are fewer monotonic
paths.

The origin of the positive correlation between the local roughness and
path divergence (Fig.\ \ref{fig:5}) is less obvious.  Paradoxically,
greater noise results in lower mean local roughness of noisy additive
landscapes.  The lowering of the overall mean fitness with noise and,
more importantly, the flattening of the mean fitness dependence on the
distance from the peak (Fig.\ \ref{fig:7}) appear to provide an
explanation for this counter-intuitive result.  Indeed we found that
in noisy additive landscapes there is a characteristic fitness value
of approximately 0.2 above which roughness increases with increasing
noise and below which roughness declines with increasing noise.  Given
that roughly 75\% of the points on the landscape have fitnesses below
0.2, the landscape-averaged local roughness declines with increasing
noise amplitude.

\section*{Discussion}
\label{sec:discussion}

Here we examined the fraction of monotonic paths and introduced mean
path divergence as quantitative measures of the degree to which the
starting and ending points determine the path of evolution on fitness
landscapes.  The lower the mean path divergence value, the more
deterministic (and predictable) evolution is.  Global measures of
landscape roughness correlate with path divergence in the three
analyzed classes of fitness landscapes: additive landscapes perturbed
by noise, landscapes derived from our protein folding model and two
small empirical landscapes.  The folding landscapes are substantially
smoother than their permuted counterparts.  As a result, although in
all analyzed landscapes only a small fraction of the theoretically
possible evolutionary trajectories is accessible, this fraction is
much greater in the folding and experimental landscapes than it is in
randomized landscapes.  In addition, the mean path divergence in the
randomized landscapes is significantly smaller than in the original
landscapes.  Thus, the model and empirical landscapes possess similar
global architectures with many more diverged monotonic paths to the
high peaks than uncorrelated landscapes with the same distribution of
fitness values.  Consequently, evolution in fitness landscapes is
substantially more robust to random mutations and less deterministic
(less predictable) than expected by chance.  These findings are
compatible with the concept that might appear counter-intuitive but is
buttressed by results of population genetic modeling, namely, that
robustness of evolving biological systems promotes their evolvability
\cite{wagner08:_robus_and_evolv,draghi10:_mutat_robus_can_facil_adapt,masel10:_robus_and_evolv}.
Additionally, the folding landscapes exhibit a substantial deficit of
peaks compared to perturbed additive landscapes and experimental
landscapes, a property that translates into a substantially greater
fraction of paths leading to the main peak.  

When it comes to the interpretation of the properties of fitness
landscapes described here, an inevitable and important question is
whether the folding model employed here is sufficiently complex and
realistic to yield biologically relevant information.  In selecting
the complexity of our folding model, we attempted to construct the
simplest model which exhibits 1) a rich spectrum of low energy
conformations across the sequence space, and 2) a non-trivial
distribution of substitutions effects on the low energy conformations.
An important choice is whether the location of monomers is confined to
a lattice or can be varied continuously.  When the configuration space
is continuous, the distribution of energy barriers between
energetically optimal conformations can extend to zero.  Therefore,
the subtlety of distinctions between conformations can lead to a
richer structure of the fitness landscape.  We chose not increase the
complexity of the model further and treated monomers as point-like
particles in a chain where the distance between nearest neighbors is
fixed but the angle between successive links in the chain in
unrestricted.  Our level of abstraction is therefore somewhere between
lattice models and all-atom descriptions of proteins
\cite{govindarajan98:_therm_hypot_prot_fold,taverna02:_mut_robust,goldstein11:_evol_marg_stab,tiana04:_evol_designability,shakhnovich06:_prot_fold_therm_dynam,zeldovich07:_protein_stabil_limit_complex,zeldovich08:_under_protein_evolut,bastolla99:_neutr_evol_latt_protein,bastolla06:_indep_sites_AA_dist,zhang09:_protein_foldin_simul}.

Another important choice is the number of the model monomer
types. Again, we opted for an intermediate level of abstraction and
chose four types of monomers: hydrophobic, hydrophilic, and
charged. This choice drastically reduces the size of the sequence
space while retaining some of the substitution complexity whereby
hydrophilic and charged monomers can be swapped under some conditions
without radically altering the native state. The intermediate level of
abstraction in our approach has its pros and cons.  Although the model
reproduces key features of protein folding such as the existence of
the hydrophobic folding nucleus and two-stage folding kinetics
\cite{gillespie04:_prot_fold_rate_theory,finkelstein07:_fold_rate_nucl_glob},
compact conformations certainly do not represent proteins. Rather, we
might think of our monomers as representing structurally grouped
regions several (perhaps up to a dozen) amino-acids in length. Compact
conformations in the model might therefore be analogous to tertiary
structures of proteins. Representing sequence space with only four
monomer types and treating mutations without reference to the
underlying DNA or genetic code does not accurately reflect the natural
mutation process. However, our goal was to isolate the features of
fitness landscapes which could be traced directly to the constraints
imposed by the heteropolymer folding kinetics and energetics. We
therefore used a simple sequence space and a homogeneous mutation
model to avoid compounding the fitness landscape structure by the
complexity derived from the mutation process.

Most importantly, our folding model has been shown to reproduce the
observed universal distribution of the evolutionary rates of
protein-coding genes as well as the dependencies of the evolutionary
rate on protein abundance and effective population sizes
\cite{lobkovsky10:_univer_distr}.  Therefore, despite its simplicity,
the behavior of this model might reflect important aspects of protein
evolution. In particular, the conclusions drawn from the analysis of
the model landscapes exhaustively explored here could also apply to
the fitness landscapes of protein evolution.  In the previous work, we
concluded that the universal distribution of evolutionary rates and
other features of protein evolution follow from the fundamental
physics of protein folding \cite{lobkovsky10:_univer_distr}.  The
results presented here suggest that the (relative) smoothness and a
substantial deficit of peaks in the fitness landscapes of protein
evolution that lead to mutational robustness and the ensuing
evolvability could similarly follow from the fact that proteins are
heteropolymers that have to fold in three dimensions to perform their
functions.

The experimental landscapes considered here are decidedly incomplete.
Due to experimental limitations, only the analysis of binary
substitutions at a handful of sites is feasible at this time.  The
incompleteness of the empirical landscapes analyzed in this work could
be the cause of the observed lack of peak suppression.  This
proposition will be put to test by the study of larger parts of
experimental landscapes that are becoming increasingly available.

\section*{Materials and Methods}
\subsection*{Folding model}
\label{sec:fold-model}

The goal of this study is to explore the relationship between
roughness and path divergence in realistic fitness landscapes.  Our
polymer folding model provides a simple way of constructing such
landscapes. The model has been described in detail previously
\cite{lobkovsky10:_univer_distr}.

In brief, the model polymer is a flexible chain of monomers in which
the the nearest neighbors interact via a stiff harmonic spring
potential with rest length $a = 1$.  The angles between the successive
links in the chain are unrestricted.  There are four types of
monomers: hydrophobic \textsf{H}, hydrophilic \textsf{P}, and charged
\textsf{+} and \textsf{--}.  Next nearest neighbors $i$ and $j$ in the
chain and beyond interact via a pairwise potential
\begin{equation}
  \label{eq:3}
  U_{ij}(r_{ij}) = \frac{A_{ij}}{r_{ij}^{12}} - \frac{C_{ij}}{r_{ij}^6} +
    \frac{q_i q_j e^{-D \, r_{ij}}}{r_{ij}},
\end{equation}
where $r_{ij}$ is the distance between monomers $i$ and $j$, $q_i$ is
the monomer's charge, $D$ is the Debye-H\"uckel screening length, and
$A_{ij}$ and $C_{ij}$ depend on the pair in question.  The interaction
parameters are chosen to mimic the essential features of the
amino-acid interactions.  To emulate the effects of solvent, we assign
a stronger attraction to the \textsf{HH} pair than to the \textsf{PP},
\textsf{++}, and \textsf{--\,--} pairs.  There is also a long range
repulsion between \textsf{H} and \textsf{P} and even stronger
repulsion between \textsf{H} and the charged monomers.  The values of
the parameters are $q_{\pm} = \pm 2$, Debye-H\"uckel screening length
$D = 3$.  The Lennard-Jones coefficients $A_{ij}$ and $C_{ij}$ are
\begin{eqnarray}
  \label{eq:6}
  && A_{HH} = 4, \ A_{HP} = A_{H+} = 2, \ A_{PP} = A_{P+} = A_{++} =
  1, \nonumber \\
  && C_{HH} = 8, \ C_{HP} = -1, \ C_{H+} = -3, \ C_{PP} = C_{P+} =
  C_{++} = 2.
\end{eqnarray}
Note that a $+$ can be substituted by a $-$ in the subscripts and the
coefficients are symmetric with respect to the interchange of the
indices.

The energy of the chain is
\begin{equation}
  \label{eq:4}
  E = \sum_{|i - j| > 1} U_{ij} + \frac{b \, T}{2} \sum_{i = 1}^{N-1}
  (r_{i,i+1} - a)^2,
\end{equation}
where the first term is the sum of the pairwise energies given by
Eq.~(\ref{eq:3}) over non-nearest neighbor pairs, and the second term
reflects the springs connecting nearest neighbors.  The spring
constant is proportional to temperature $T$.  The parameters are fixed
for all simulation runs at $b = 300$, and the quench temperature $T =
1$.  To mimic the observed tendency of the $N$ and $C$ termini to be
in close proximity, we fixed the endpoint monomers of the model
sequences to be of $+$ and $-$ types.

Dynamics of folding are simulated via overdamped Brownian kinetics
which are appropriate when inertial and hydrodynamic effects are not
important.  Units are chosen so that each component $\alpha$ of the
$i$'th monomer's coordinates $x_{\alpha i}$ is updated according to
\begin{equation}
  \label{eq:5}
  x_{\alpha i}(t + \Delta t) = x_{\alpha i}(t) - \frac{\Delta
    t}{T} \, \frac{\partial E}{\partial x_{\alpha i}}(t) + 
  W_{\alpha i}(t),
\end{equation}
where $\Delta t$ is the time step and $W_{\alpha i}(t)$ is a random
variable with zero mean, variance $2 \Delta t$, uncorrelated with $W$
for other times, monomers and spatial directions.

\subsection*{Native structure ensemble and correct folding probability}
\label{sec:native-struct-ensemb}

The ``native structure'' of a particular sequence is represented by an
equilibrium ensemble of conformations.  The ensemble is constructed by
identifying the typical folded conformation and measuring the
characteristic RMSD $D$ due to thermal fluctuations in the folded
state.  Three thousand quenches are then performed and the resulting
folded conformations are accumulated.  The equilibrium ensemble that
represents the native structure is defined as the largest cluster of
quenched conformations within RMSD distance $D$ from each other.
Thus, each conformation in the ensemble differs from any other by an
amount comparable to the differences introduced by thermal
fluctuations alone.

The concept of the native structure ensemble allows us to compute the
probability that a sequence folds to a particular structure in a
natural, physically plausible fashion.  Given a native structure
ensemble we assess its conformation space density by computing the
distance $d_i$ between each member $i$ of the ensemble and its closest
neighbor.  Given the set $\{d_i\}$ of these shortest distances we
compute the median $Q$ and the median absolute deviation (MAD) $V$.  A
new conformation is deemed to belong to the ensemble if the shortest
distance from this conformation to the members of the ensemble is
smaller than $R = Q + 3V$.

Given a native structure ensemble of some sequence $s_1$ we compute
the probability $P$ that sequence $s_2$ (which could be $s_1$ itself)
folds to the this structure by accumulating $M = 100$ equilibrated
quenched conformations of $s_2$ and using the above criterion to
determine the fraction $P$ that belong to the native structure
ensemble of $s_1$.  Because $M = 100$ sample conformations are
computed, the smallest measurable $P$ is $1/M = 0.01$.  The sample
size used to measure $P,$ dictated by the computational demands of the
model, introduces a random component to the model fitness landscapes.
As we report below, model landscapes turn out to be substantially
smoother than random.  Therefore the underlying global structure of
the model landscapes appears to survive the modest amount of
randomness introduced by the relatively small sample size used for
measuring $P$.

\subsection*{Search for compact robust folders}
\label{sec:search-comp-robust}

Robust folders (sequences with a high probability of correct folding)
tend to have large linear regions stretched by repulsive Coulomb
interactions.  Because the linear regions have no contacts with other
monomers, we focused our attention on compact conformations with a
high monomer contact density.  Substitutions in these higher
complexity conformations were more likely to exhibit non-trivial
effects.  To find compact robust folders in the vast available
sequence space of $23$-mers (the sequences are of length $N=25$ but
the endpoint monomer types are fixed) with $4$ monomer types, we
implemented a simulated annealing search which optimized the correct
folding probability $P$ divided by the cube of the native
conformation's radius of gyration.  The search produced over 800
sequences with $P > 0.5$ and at least two distinct regions of the
polymer in mutual contact.

\subsection*{Assembly of the folding fitness landscapes}
\label{sec:assembly-fitn-landsc}

We examined each single substitution mutant of a robustly folding
sequence and computed the folding probability $P$ to the structure of
the original sequence.  All mutants with $P > 0$ were added to the
landscape and if $P\ge 0.1$ their mutants were also examined.  This
process is repeated until all mutants of the last sequence under
consideration have $P < 0.1$.

From our study of complete landscapes we estimate that on average for
each sequence with $P > 0$ which is included into the landscape,
roughly 6 others with $P = 0$ need to be examined.  Since each quench
and equilibration takes about 2--4 seconds, landscape construction
takes roughly 30 minutes to an hour per included sequence.  Thus
landscapes larger than 10,000 sequences take months to compile.

At the time of submission, 39 complete landscapes have been
constructed, the largest comprising 12969 sequences.

\subsection*{Additive landscapes perturbed by noise}
\label{sec:addit-landsc-pert}

The organization of the folding fitness landscapes and experimental
landscapes were compared with perfectly additive landscapes perturbed
by noise constructed as follows.  Each substitution to the peak
fitness sequence was assigned a negative fitness differential drawn at
random from an exponential distribution with parameter $\lambda = 3$.
The sum over the fitness differentials of a particular set of
substitution was modified by either additive of multiplicative noise
\cite{aita00:_Fuji_noise}.  Additive noise is drawn from a Gaussian
distribution with zero mean and standard deviation $\nu$ which was
varied between $0.05$ and $0.5$.  The multiplicative perturbation is
achieved by multiplying the fitness by a number drawn from a uniform
distribution $[0,1)$ raised to a positive power $\mu$ varied between
$0.1$ and $10.$ When $\mu$ is small, multiplicative factors are close
to unity and the perturbation is small as well.  If the perturbed
fitness was positive, the mutant was included into the landscape.  The
noise amplitude was varied to obtain a family of landscapes of
continuously varying roughness.  Only the data for the additive
landscapes with multiplicative noise were included in this manuscript.
Landscapes perturbed by other types of noise exhibited essentially the
same qualitative behavior.

\subsection*{Experimental landscapes}
\label{sec:exper-landsc}

The studies on experimental fitness landscapes typically involve
constructing a library of all possible combinations of binary
mutations at a small number of sites. The first study included in the
present analysis measured the minimum inhibitory concentrations (MIC)
of an antibiotic for a complete spectrum of mutants with modified TEM
$\beta$-lactamases; the transition from the antibiotic-sensitive to
the antibiotic-resistant form requires five mutation, so the landscape
encompassed 120 mutational trajectories between the most distant
points on the landscape (or 32 sequences)
\cite{weinreich06:_5point_beta_lactam}.  The logarithm of MIC was used
as the proxy for fitness. In the second study, catalytic activity of
419 sesquiterpene synthase mutants that differed by at most 9
substitutions was measured \cite{omaille08:_sesquiterpene}.  We used
the catalytic specificity (propensity for producing a particular
reaction product rather than a broad spectrum of products) of the
mutant enzymes as the proxy for fitness.  Before performing the
analysis, the fitnesses in the experimental landscapes are mapped onto
the $[0.01,1)$ interval to enable meaningful quantitative comparisons
of the roughness measures.



\newpage
\section*{Figure Legends}

\begin{figure}[!ht]
  \centering
  \caption{Roughness, monotonic paths and suboptimal peak suppression
    in folding and experimental landscapes.  (A) Deviation from
    additivity for the folding landscapes (larger symbols), their
    scrambled versions (smaller symbols) and the two experimental
    landscapes.  Error bars show one standard deviation within the
    ensemble of permuted landscapes.  (B) Fraction of monotonic paths
    to the main peak in folding, scrambled and experimental
    landscapes.  (C) The number of peaks is vastly greater in
    scrambled landscapes than in folding or experimental landscapes
    (with the exception of the sesquiterpene synthase landscape).}
  \label{fig:1}
\end{figure}

\begin{figure}[!ht]
  \centering
  \caption{The Z-scores of different characteristics of the original
    folding and experimental landscapes measured with respect to the
    ensembles of their randomly permuted counterparts.}
  \label{fig:2}
\end{figure}

\begin{figure}[!ht]
  \centering
  \caption{Correlations between different quantitative characteristics
    of the folding landscapes.  Each panel quotes the Spearman rank
    correlation coefficient between the particular pair of
    characteristics.}
  \label{fig:3}
\end{figure}

\begin{figure}[!ht]
  \centering
  \caption{Mean path divergence as a function of selection pressure
    for a folding landscape with 5936 nodes and 65 peaks.  Solid lines
    are labeled by the Hamming distance between the pairs of starting
    and ending points of the trajectory bundles over which the path
    divergence is averaged.}
  \label{fig:4}
\end{figure}

\begin{figure}[!ht]
  \centering
  \caption{The dependence of the path divergence (top row) and the
    monotonic path fraction (bottom row) on the measures of landscape
    roughness.  The dots of different color correspond to noisy
    additive landscapes with differing amounts of multiplicative
    noise: low (red), two intermediate levels (green smaller than
    blue), and high (magenta).  Yellow circles represent the folding
    landscapes, the cyan squares--the $\beta$-lactamase landscape, and
    the red triangles--the sesquiterpene synthase landscape.}
  \label{fig:5}
\end{figure}

\begin{figure}[!ht]
  \centering
  \caption{Mean path divergence in folding and experimental landscapes
    (larger symbols) landscapes, as well as their scrambled versions
    (smaller symbols) as a function of Hamming distance from the main
    peak.}
  \label{fig:6}
\end{figure}

\begin{figure}[!ht]
  \centering
  \caption{Fitness averaged over all points at a particular distance
    $H$ from the peak for folding landscapes, additive landscapes with
    the same three levels of multiplicative noise used in Fig.\
    \ref{fig:5} and the sesquiterpene synthase landscape.}
  \label{fig:7}
\end{figure}

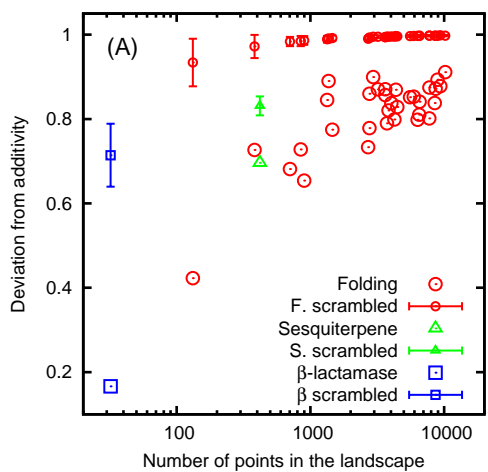
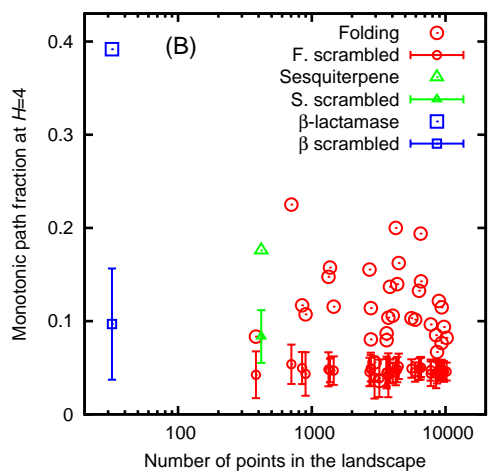
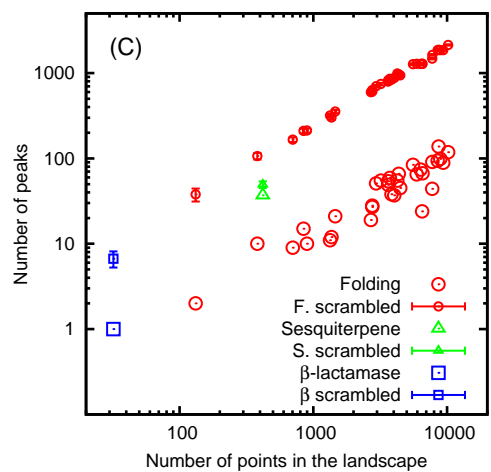
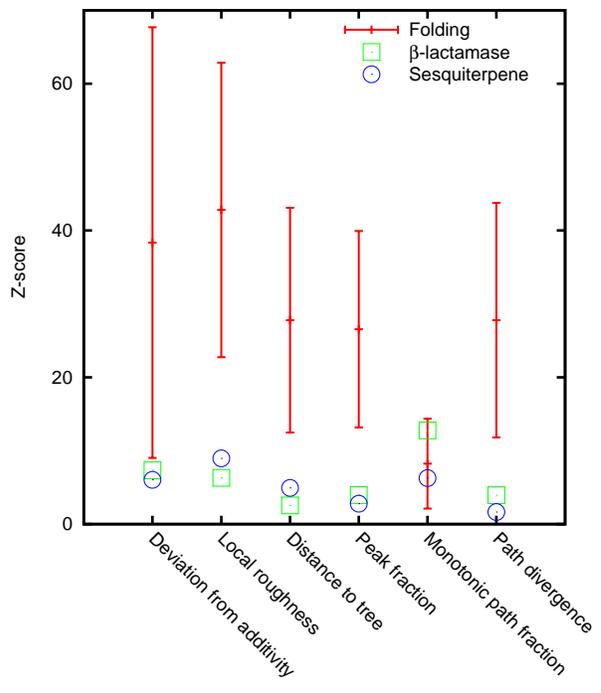
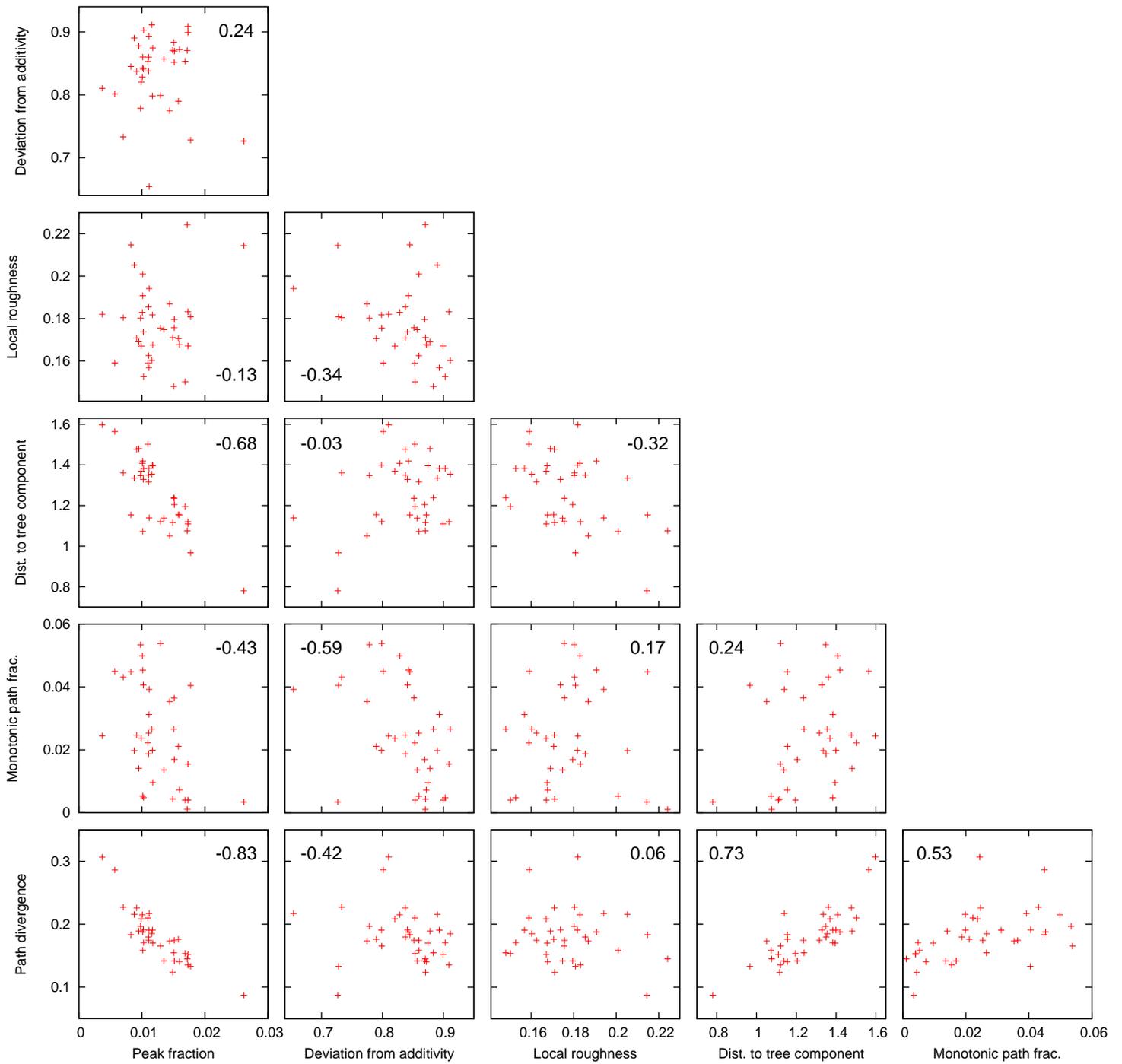
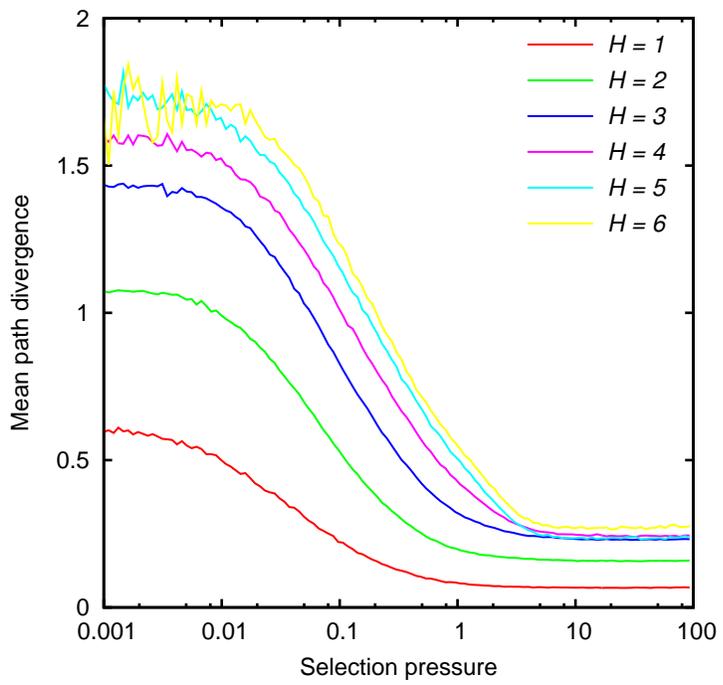
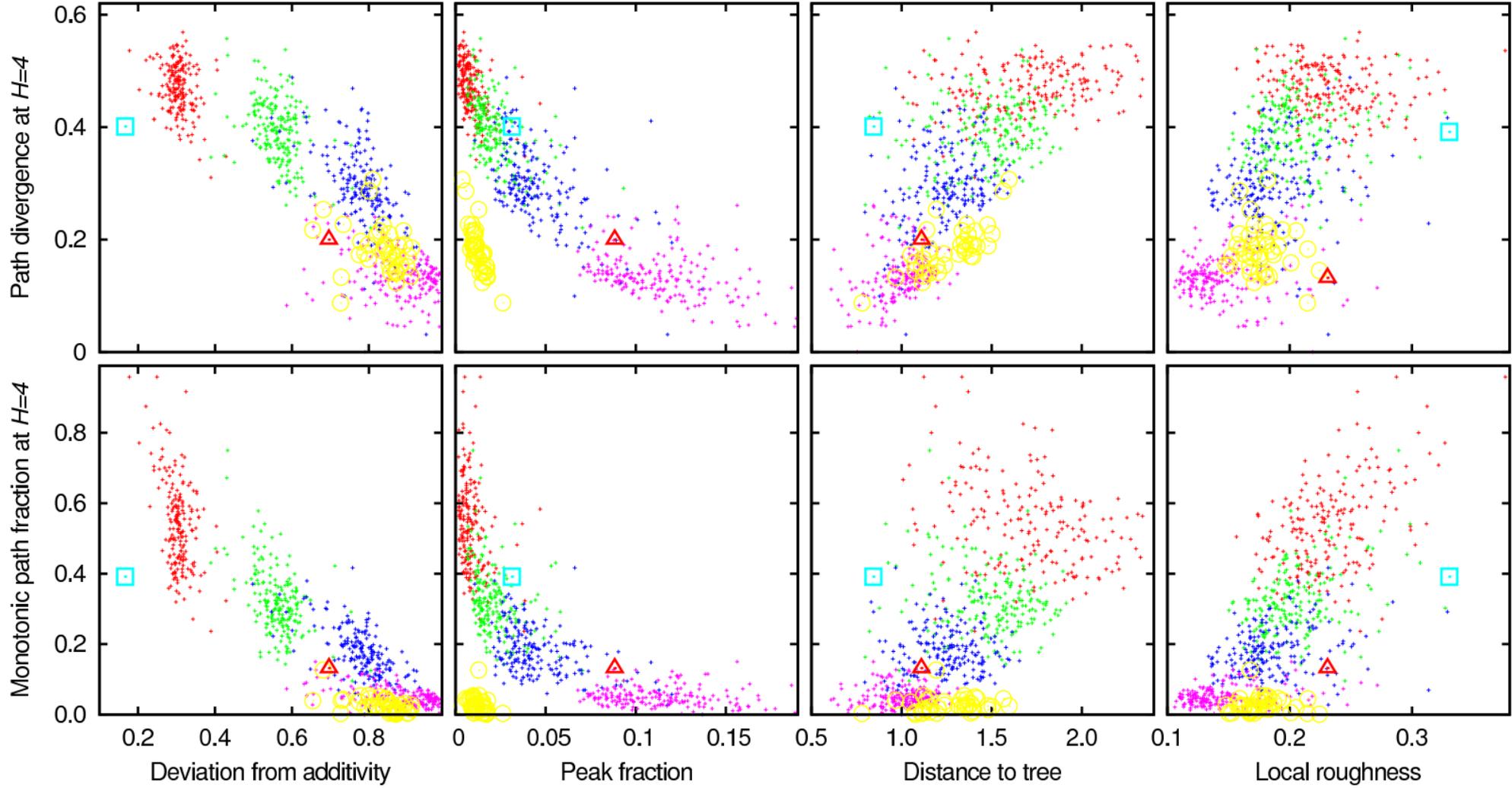
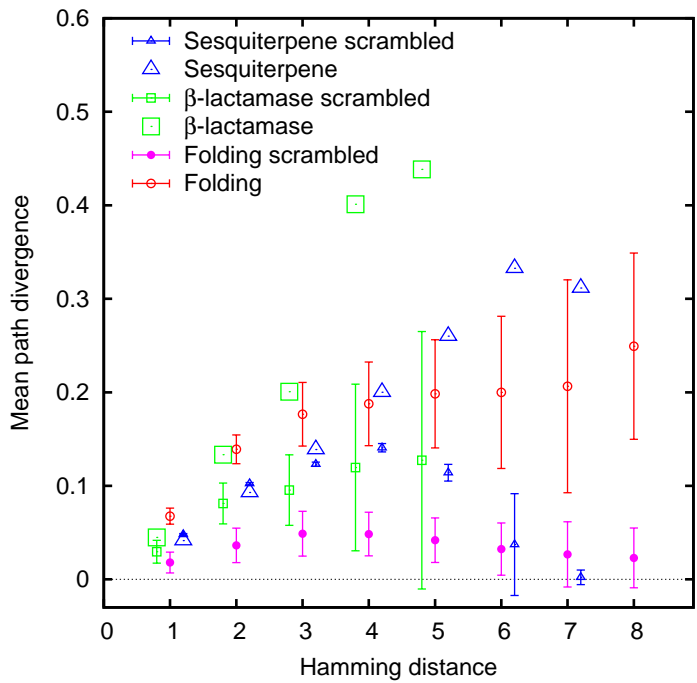
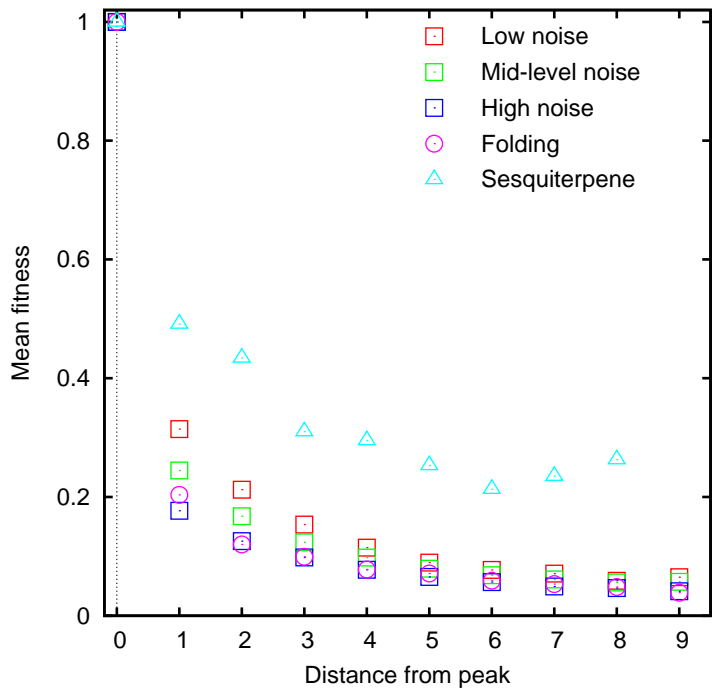
\end{document}